\documentclass[amsmath,amssymb,prb,preprintnumbers,superscriptaddress, twocolumn,showpacs]{revtex4}

\usepackage{bm}
\usepackage{amssymb}
\usepackage{amsmath}
\usepackage{graphics}
\usepackage{epsfig}
\usepackage[usenames]{color}
\usepackage{ulem}

\begin{document}

\title{Nb nano superconducting quantum interference devices with high spin sensitivity for operation in magnetic fields up to 0.5\,T}

\author{R.~W\"{o}lbing}
\author{J.~Nagel}
\author{T.~Schwarz}
\affiliation{%
  Physikalisches Institut -- Experimentalphysik II and Center for Collective Quantum Phenomena in LISA$^+$,
  Universit\"at T\"ubingen,
  Auf der Morgenstelle 14,
  D-72076 T\"ubingen, Germany
}

\author{O.~Kieler}
\author{T.~Weimann}
\author{J.~Kohlmann}
\author{A.B.~Zorin}
\affiliation{%
  Fachbereich 2.4 "Quantenelektronik", Physikalisch-Technische Bundesanstalt,
  Bundesallee 100,
  D-38116 Braunschweig, Germany
}

\author{M.~Kemmler}
\author{R.~Kleiner}
\author{D.~Koelle}
\affiliation{%
  Physikalisches Institut -- Experimentalphysik II and Center for Collective Quantum Phenomena in LISA$^+$,
  Universit\"at T\"ubingen,
  Auf der Morgenstelle 14,
  D-72076 T\"ubingen, Germany
}

\date{\today}

\begin{abstract}

We investigate electric transport and noise properties of microstrip-type submicron direct current superconducting quantum interference devices (dc SQUIDs) based on Nb thin films and overdamped Josephson junctions with a HfTi barrier.
The SQUIDs were designed for optimal spin sensitivity $S_\mu^{1/2}$ upon operation in intermediate magnetic fields $B$ (tens of mT), applied perpendicular to the substrate plane.
Our so far best SQUID can be continuously operated in fields up to $B\approx\pm50\,\rm{mT}$ with rms flux noise $S_{\Phi,\rm w}^{1/2}\leq250\,\rm{n\Phi_0/Hz^{1/2}}$ in the white noise regime and spin sensitivity $S_{\mu}^{1/2}\leq29\,\rm{\mu_B/Hz^{1/2}}$.
Furthermore, we demonstrate operation in $B=0.5\,\rm{T}$ with high sensitivity in flux $S_{\Phi,\rm w}^{1/2}\approx680\,\rm{n\Phi_0/Hz^{1/2}}$ and in electron spin $S_{\mu}^{1/2}\approx79\,\rm{\mu_B/Hz^{1/2}}$.
We discuss strategies to further improve the nanoSQUID performance.

\end{abstract}


\maketitle

Recent developments in miniaturized submicron-sized direct current (dc) superconducting quantum interference devices (SQUIDs) are motivated by the need of sensitive detectors for small spin systems such as molecular magnets\cite{Gatteschi03,Wernsdorfer07,Bogani08a} and magnetic nanoparticles,\cite{Wernsdorfer01} cold atom clouds\cite{Fortagh05} or single electrons and atoms\cite{Bushev11} and improved resolution in scanning SQUID microscopy.\cite{Kirtley95,Black95,Veauvy02,Kirtley09,Finkler10,Finkler12}
As a common approach, nanoSQUIDs based on constriction Josephson junctions (JJs) have been used,\cite{Foley09,Faucher09,Chen10a,Romans11,Russo12} achieving root mean square (rms) flux noise power $S_\Phi^{1/2}$ down to a few $100\,\rm{n\Phi_0/Hz^{1/2}}$ ($\Phi_0$ is the magnetic flux quantum) in magnetically shielded environment.\cite{Hao08}
However, constriction JJs, even if resistively shunted, often show hysteretic current-voltage-characteristics (IVCs).
This hampers continuous SQUID operation as required for the investigation of magnetization dynamics of magnetic particles and the use of common SQUID electronics, developed for readout of very sensitive dc SQUIDs with nonhysteretic JJs.
Furthermore, the noise properties of constriction JJs are not well understood, which makes SQUID optimization difficult.

An alternative approach is the use of submicron superconductor-normal conductor-superconductor (SNS) sandwich-type JJs, which offer large critical current densities in the $10^5\,\rm{A/cm^2}$ range and which are intrinsically shunted, providing nonhysteretic IVCs without the need of bulky external shunt resistors.\cite{Nagel11a}
In a standard thin film SQUID geometry, the SQUID loop and the JJ barrier are in the plane of the thin films.
For detection of magnetization reversal of a small magnetic particle, one applies an external magnetic field in the plane of the SQUID loop and detects the change of the stray field coupled to the SQUID upon magnetization reversal, without coupling the external field to the SQUID.
However, in this case, the applied field also couples magnetic flux into the JJ barrier and reduces its critical current, which in turn reduces the SQUID sensitivity.
In order to avoid this problem, in this letter we present results on a modified SQUID design, which takes advantage of the multilayer technology used for SNS JJ fabrication.
This approach allows for a further reduction of the SQUID inductance and hence improved SQUID sensitivity and at the same time operation in higher magnetic fields.

\begin{figure}[t]
\includegraphics[width=7cm]{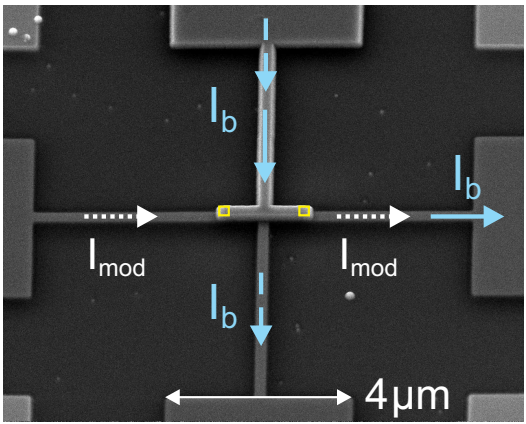}
\caption{(Color online) Scanning electron microscopy (SEM) image of SQUID\,2.
Open (yellow) squares indicate positions of JJs.
Arrows indicate current paths for bias current $I_{\rm{b}}$ (dashed: symmetric bias; solid: asymmetric bias) and modulation current $I_{\rm{mod}}$ (dotted).}
\label{Fig:Figure1}
\end{figure}

The Nb thin film dc SQUIDs have a microstrip geometry, i.e.~the two $250\,\rm{nm}$ wide arms of the SQUID loop lie directly on top of each other.
The $200\,\rm{nm}$ thick bottom and $160\,\rm{nm}$ thick top Nb layers are separated by a $225\,\rm{nm}$ thick insulating $\rm{SiO_2}$ layer and are connected via two JJs with areas $200\times200\,\rm{nm^2}$ and a nominally $24\,\rm{nm}$ thick HfTi barrier (see Fig.~\ref{Fig:Figure1}).
HfTi was chosen as a barrier material as, among other binary materials, it provides a relatively high resistivity, does not become superconducting at 4.2\,K and is compatible with our fabrication technology.
For details on sample fabrication and JJ properties we refer to [\onlinecite{Hagedorn02, Hagedorn06, Nagel11a}].
The size of the SQUID loop is defined by the $1.6\,\rm{\mu m}$ spacing between the JJs and by the $\rm{SiO_2}$ interlayer thickness.
In contrast to earlier work,\cite{Nagel11a} for this geometry a sufficiently large magnetic field $B$ can be applied perpendicular to the substrate plane without inducing a significant magnetic flux penetrating either the SQUID loop or the junction barrier.
Furthermore, this design provides a very small area of the SQUID loop and hence a very small SQUID inductance $L$ of a few pH or even lower.
This is essential for reaching ultra-low values for the spectral density of flux noise power $S_\Phi$.\cite{VanHarlingen82}

For current and flux biasing, additional $250\,\rm{nm}$ wide Nb lines are connecting the SQUID in a cross-shape geometry, and a bias current $I_{\rm{b}}$, flowing from the top Nb layer through the JJs to the bottom Nb layer, can be applied either in a symmetric or asymmetric configuration; see Fig.~\ref{Fig:Figure1}.
For simplified readout we use asymmetric current bias in the following.
A magnetic flux $\Phi$ can be coupled into the SQUID loop by applying a modulation current $I_{\rm{mod}}$ across the bottom Nb layer ("flux bias line").
This enables flux biasing the SQUIDs at the optimum working point without the need of an external coil.
Furthermore, the flux bias line can also be used to provide a feedback flux for SQUID operation in a flux locked loop.
However, in this work, the SQUIDs were always read out open loop.

We investigated various SQUIDs which were fabricated in two different runs on separate wafers.
Below we present results for two devices, SQUID\,1 from wafer\,1 and SQUID\,2 from wafer\,2.
The main difference in the design of these devices is the different lengths $\sim 2.5\,\rm{\mu m}$ (SQUID\,1) and $\sim 5\,\rm{\mu m}$ (SQUID\,2) of the narrow bias lines, running from the center of the SQUID to the $4\,\rm{\mu m}$ wide connection lines further away from the SQUID (cf.~Fig.~\ref{Fig:Figure1}).
This variation has a strong impact on the SQUID performance in applied magnetic fields, as will be shown below.


\begin{figure}[b]
\includegraphics[width=8.5cm]{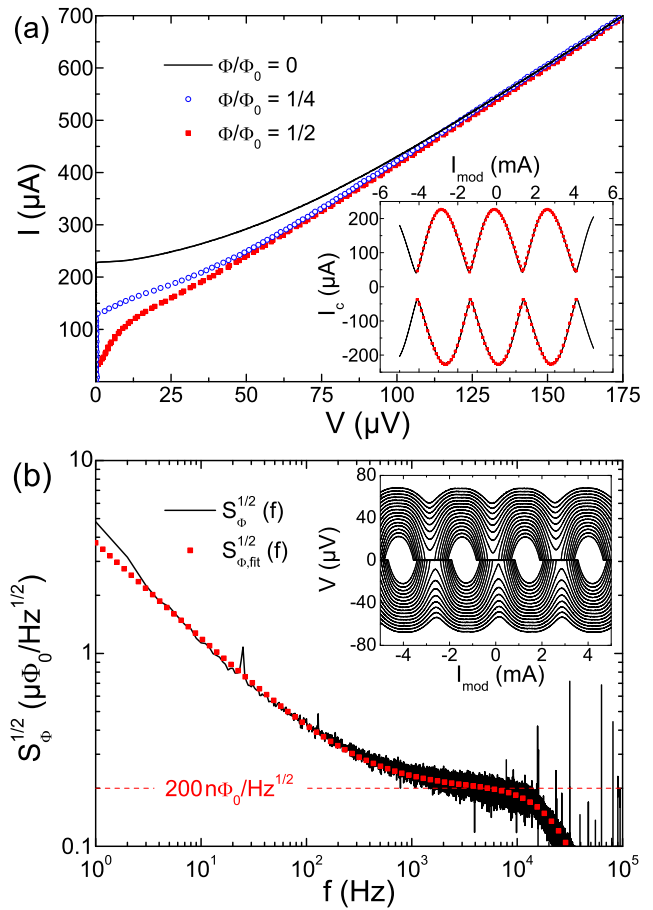}
\caption{(Color online) Transport and noise characteristics of SQUID\,2.
(a) IVCs for different flux $\Phi$; inset shows measurement (solid black lines) and numerical simulation (dotted red lines) of $I_{\rm{c}}(I_{\rm{mod}})$.
(b) Solid black line: Spectral density of rms flux noise $S_\Phi^{1/2}(f)$ at optimum working point ($I_{\rm{b}} = 230\,\rm{\mu A}$, $I_{\rm{mod}}=243\,\rm{\mu A}$).
Dotted (red) line: fitted spectrum; dashed (red) line indicates white noise level for fitted curve.
Inset shows $V(I_{\rm{mod}})$ for $I_{\rm{b}} =\pm(150\ldots300)\,\rm{\mu A}$ (in $10\,\rm{\mu A}$ steps).}
\label{Fig:Figure2}
\end{figure}


All data were taken at temperature $T=4.2\,\rm{K}$.
We first present results of transport and noise measurements of the two SQUIDs in an electrically and magnetically shielded environment.
Since both devices showed qualitatively the same behavior, we only give a detailed analysis of SQUID\,2 and summarize the main parameters extracted for both devices in Table \ref{Tab:SQUIDs_parameters}.
Regarding absolute values, a major difference between both devices are the values for maximum critical current $I_{\rm{c}}$ and normal resistance $R_{\rm{N}}$, which probably is due to variations in the HfTi barrier thickness for the devices fabricated in different runs.
Devices fabricated from the same run showed a spread in $I_{\rm{c}}$ and $R_{\rm{N}}$ values of $\pm 10\,\%$.

Figure \ref{Fig:Figure2}(a) shows IVCs of SQUID\,2 for $\Phi/\Phi_0=0$, 1/4 and 1/2.
The IVCs are nonhysteretic with $I_{\rm{c}}=227\,\rm{\mu A}$ and $R_{\rm{N}}=250\,\rm{m \Omega}$, yielding a characteristic voltage $V_{\rm{c}} \equiv I_{\rm{c}} R_{\rm{N}}=57\,\rm{\mu V}$.
The IVC at $\Phi/\Phi_0=1/2$ exhibits a small bump for low voltages.
This bump appears in all our devices and is presumably a property of the quasiparticle current rather than a $LC$ resonance of the SQUID.
The inset of Fig.~\ref{Fig:Figure2}(a) shows the modulation $I_{\rm{c}}(I_{\rm{mod}})$ for positive and negative bias current.
From the modulation period we obtain the inverse mutual inductance $M_{\rm{i}}^{-1}=2.73\,\rm{mA/\Phi_0}$.
From the modulation depth we find a screening parameter $\beta_{\rm{L}} \equiv 2I_0L/\Phi_0=0.25$.
By assuming that both JJs are identical, i.e.~$I_{\rm{c}} \equiv 2I_0$, we determine the SQUID inductance $L=2.3\,\rm{pH}$.

The $V(I_{\rm{mod}})$ modulation for different bias currents, plotted in the inset of Fig.~\ref{Fig:Figure2}(b), yields a maximum transfer function $V_{\Phi}\equiv{\partial V}/{\partial \Phi}=164\,\rm{\mu V/\Phi_0}$ for $I_{\rm{b}}=230\,\rm{\mu A}$.
The shift in $I_{\rm{c}}(I_{\rm{mod}})$ and $V(I_{\rm{mod}})$ for positive and negative bias currents can be attributed to the asymmetric current bias, which leads to an inductance asymmetry $\alpha_{\rm{L}}\equiv(L_2-L_1)/(L_1+L_2)$; here $L_1$ and $L_2$ are the inductances of the two SQUID arms.
The measured $I_{\rm{c}}(I_{\rm{mod}})$-characteristics are fitted well by numerical simulations based on coupled Langevin equations\cite{Chesca-SHB-2} with a noise parameter $\Gamma \equiv 2 \pi k_{\rm{B}}T/I_0 \Phi_0=1.55\cdot10^{-3}$ ($k_{\rm{B}}$ is the Boltzmann constant) and $\alpha_{\rm{L}}=-0.35$ (see inset of Fig.~\ref{Fig:Figure2}(a); dotted lines).

\renewcommand{\arraystretch}{1.5}
\begin{table}[t]
\caption{Parameters of SQUID\,1 and SQUID\,2.}
\begin{center}
\begin{tabular}{ccccccccc}
\hline\hline
        & $I_{\rm{c}}$      & $R_{\rm{N}}$      & $I_{\rm{c}}R_{\rm{N}}$    & $\beta_{\rm{L}}$  & $L$   & $M_{\rm{i}}^{-1}$             & $V_\Phi$                      & $S_{\Phi,\rm{w}}^{1/2}$\\
        & ($\rm{\mu A}$)    & ($\rm{m\Omega}$)  & ($\rm{\mu V}$)            &                   & (pH)  & $(\frac{\rm{mA}}{\Phi_0})$    & $(\frac{\mu\rm{V}}{\Phi_0})$  & $(\frac{\rm{n}\Phi_0}{\rm{Hz^{1/2}}})$\\
\hline
SQUID\,1& 129               & 385               & 50                        & 0.19              & 3.0   & 2.63                          & 154                           & 260\\
SQUID\,2& 227               & 250               & 57                        & 0.25              & 2.3   & 2.73                          & 164                           & 200\\
\hline\hline
\end{tabular}
\end{center}
\label{Tab:SQUIDs_parameters}
\end{table}

Using a commercial SQUID amplifier with a voltage noise $S_{\rm{V}}^{1/2} \approx 40\,\rm{pV/Hz^{1/2}}$ and a -3\,dB cutoff frequency $f_{\rm{c}} \approx 30\,\rm{kHz}$, we measured the spectral density of the rms flux noise $S_{\Phi}^{1/2}(f) \equiv S_{\rm{V}}^{1/2}(f)/\left|V_{\Phi}\right|$ at the optimum working point (see solid line in Fig.~\ref{Fig:Figure2}(b)).
Here the SQUID amplifier contribution was subtracted.
We observe a significant low-frequency excess noise, which wie assign to $I_0$ fluctuations in the JJs.
Since the low-frequency excess noise extends to well above $1\,\rm{kHz}$ and due to the limited bandwidth of the SQUID amplifier, we do not see a clear white noise region in the spectrum.
By fitting the experimental data (dotted line in Fig.~\ref{Fig:Figure2}(b)), we derive a low-frequency noise contribution $S_{\Phi,f}^{1/2}\propto1/f^{\alpha}$ with $\alpha=0.5$ and $S_{\Phi,f}^{1/2}(f=1\,\rm{Hz})=3.7\,\rm{\mu \Phi_0/Hz^{1/2}}$ and a white noise contribution $S_{\Phi,\rm{w}}^{1/2} = 200\,\rm{n \Phi_0/Hz^{1/2}}$ (dashed line in Fig.~\ref{Fig:Figure2}(b)).

In order to determine the spin sensitivity $S_\mu^{1/2} \equiv S_\Phi^{1/2}/\phi_\mu$ of our SQUIDs, we calculated the coupling factor $\phi_\mu$, using a routine based on the numerical solution of the London equations for the given SQUID geometry.\cite{Nagel11}
Here, $\phi_\mu\equiv\Phi/\mu$ is the magnetic flux $\Phi$ per magnetic moment $\left|\vec{\mu}\right|\equiv\mu$ coupled by a magnetic particle to the SQUID loop.
Very recently, the validity of this approach has been verified experimentally by measuring the magnetic coupling of a Ni nanotube to a Nb nanoSQUID which had the same geometry as SQUID\,2.\cite{Nagel13}
For a point-like magnetic particle with $\vec{\mu}$ perpendicular to the substrate plane, placed at a lateral distance of $10\,\rm{nm}$ from the lower edge of the upper Nb SQUID arm at the center of the loop, we obtain $\phi_\mu=8.6\,\rm{n\Phi_0/\mu_B}$ ($\mu_{\rm{B}}$ is the Bohr magneton).
Along with the obtained value of the rms flux noise $S_{\Phi,\rm{w}}^{1/2} = 200\,\rm{n \Phi_0/Hz^{1/2}}$ we calculate the spin sensitivity to $S_\mu^{1/2}=23\,\rm{\mu_B/Hz^{1/2}}$.


\begin{figure}[b]
\includegraphics[width=8.5cm]{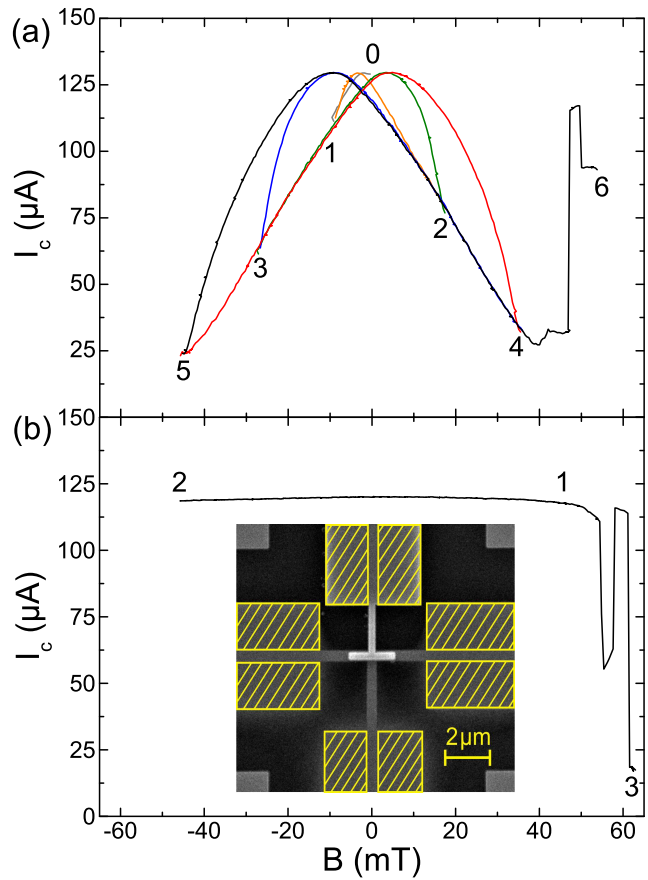}
\caption{(Color online) $I_{\rm{c}}(B)$ data of SQUID\,1 for field sweep sequence 0--6 (a) and 1--3 (b) after  removing Nb areas by FIB milling as indicated by hatched (yellow) rectangles in the inset (SEM image).}
\label{Fig:Figure3}
\end{figure}


To investigate the SQUID performance in a magnetic field $B$ applied perpendicular to the substrate plane we mounted SQUID\,1 on a high-precision alignment system (one rotator, two goniometers).
$B$ is generated by a superconducting split coil running in persistent mode to suppress field noise.\cite{Schwarz13}
Figure \ref{Fig:Figure3}(a) shows $I_{\rm{c}}(B)$ for SQUID\,1 after the alignment process for a field sweep sequence as indicated by labels 0--6.
The observed hysteresis for $|B|<45\,\rm{mT}$ is ascribed to entry and trapping of Abrikosov vortices in the $4\,\rm{\mu m}$ wide connection lines, cf.~inset of Fig.~\ref{Fig:Figure3}(b).
The steep jump in $I_{\rm{c}}$ at $B\approx45\,\rm{mT}$ can be assigned to a vortex entering the narrow Nb leads very close to the SQUID loop, as confirmed recently by magnetic force microscopy on a similar Nb nanoSQUID (with layout of SQUID\,2).\cite{Nagel13}
Subsequently, we reduced the linewidth of the connection lines of SQUID\,1 from $4\,\rm{\mu m}$ to $\sim 500\,\rm{nm}$ by focused ion beam (FIB) milling,\cite{Schwarz13} see inset of Fig.~\ref{Fig:Figure3}(b).
For the repatterned device, the maximum $I_{\rm{c}}$ was reduced by $\sim10\%$, probably due to a slight degradation of the JJs during FIB milling.
More importantly, $I_{\rm{c}}$ became almost independent of $B$, and within $B\approx\pm50\,\rm{mT}$ the magnetic hysteresis disappeared, cf.~Fig.~\ref{Fig:Figure3}(b).
At $B\approx50\,\rm{mT}$ we still observed the jump in $I_{\rm{c}}$ due to vortex entry in the narrow Nb line close to the SQUID.
This indicates that the linewidth of the Nb wiring close to the SQUID may limit the range of operation to $|B| \leq 50\,\rm{mT}$.
However, as will be shown below, even after vortex entry, by proper realignment of the applied magnetic field direction, which compensates the stray magnetic flux induced by trapped vortices, $I_{\rm{c}}$ can be restored and low flux noise can be retained.


\begin{figure}[b]
\includegraphics[width=8.5cm]{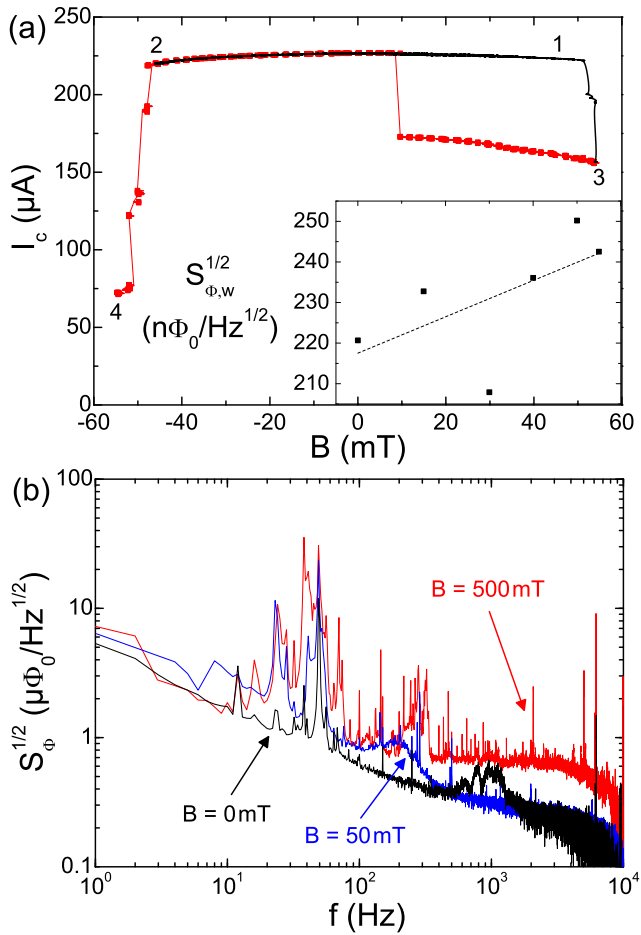}
\caption{(Color online) (a) $I_{\rm{c}}(B)$ data of SQUID\,2 for field sweep sequence 1--3 (black solid line) and 3--4 (red line plus symbols).
Inset: $S_{\Phi,\rm{w}}^{1/2}(B)$ at optimum working point; dashed line is a linear fit.
(b) Spectral density of rms flux noise $S_{\rm{\Phi}}^{1/2}(f)$ for $B=0\,\rm{mT}$, $50\,\rm{mT}$ and $500\,\rm{mT}$.}
\label{Fig:Figure4}
\end{figure}


We now turn to SQUID\,2, which has much longer narrow bias lines.
Figure \ref{Fig:Figure4}(a) shows $I_{\rm{c}}(B)$ for a field sweep $46\,\rm{mT} \rightarrow -46\,\rm{mT} \rightarrow 55\,\rm{mT}$ (1--3).
Again $I_{\rm{c}}$ is almost independent of $B$ for $|B|\leq 50\,\rm{mT}$ and, as before, we find a jump in $I_{\rm{c}}$ at $B \approx 50\,\rm{mT}$ due to a vortex entering the narrow bias lines.
The vortex can be removed by sweeping back the field as indicated by the curve (3--4) in Fig.~\ref{Fig:Figure4}(a).

For SQUID\,2 we performed noise measurements as described above to determine $S_{\Phi,\rm{w}}^{1/2}$ at several values of $B$ from $0$ to $50\,\rm{mT}$, without any jump in $I_{\rm{c}}$, see inset of Fig.~\ref{Fig:Figure4}(a).
For $B=0$, $S_{\Phi,\rm{w}}^{1/2}\approx 220\,\rm{n \Phi_0/Hz^{1/2}}$, which is slightly higher than the value obtained in the low-field setup.
We attribute this to external disturbances from the unshielded environment in the high-field setup (cf.~noise spectrum in Fig.~\ref{Fig:Figure4}(b), black line).
As indicated in the inset of Fig.~\ref{Fig:Figure4}(a), the white noise level increases only slightly with $B$ to $S_{\Phi,\rm{w}}^{1/2}\approx250\,\rm{n \Phi_0/Hz^{1/2}}$ at $B=50\,\rm{mT}$ (cf.~noise spectrum in Fig.~\ref{Fig:Figure4}(b)), still corresponding to a very small spin sensitivity $S_{\mu}^{1/2}\approx 29\,\rm{\mu_B/Hz^{1/2}}$ (in the white noise regime).
We assign this behavior to a minor decrease of $I_{\rm{c}}$ due to an imperfect alignment of the device relative to $B$.
At $B=55\,\rm{mT}$, i.e.~after the jump in $I_{\rm{c}}$ occurred and after realigning the SQUID by maximizing $I_{\rm{c}}$, we obtain a similar value $S_{\Phi,\rm{w}}^{1/2}\approx240\,\rm{n \Phi_0/Hz^{1/2}}$ as for $B=50\,\rm{mT}$.
Following the same procedure of realignment, we were able to operate the SQUID in magnetic fields up to $B=0.5\,\rm{T}$, yielding the noise spectrum as shown in Fig.~\ref{Fig:Figure4}(b), with $S_{\Phi,\rm{w}}^{1/2}\approx680\,\rm{n \Phi_0/Hz^{1/2}}$, corresponding to $S_{\mu}^{1/2}\approx 79\,\rm{\mu_B/Hz^{1/2}}$.
Note that all spectra feature excess low-frequency noise peaks, which are presumably due to mechanical vibrations of the setup.

In conclusion, we fabricated and investigated Nb nanoSQUIDs based on a trilayer geometry which were optimized for stable operation in comparatively large magnetic fields.
Very low white flux noise values down to $S_{\Phi,\rm{w}}^{1/2} \approx 200\,\rm{n \Phi_0/Hz^{1/2}}$ have been achieved in a shielded environment yielding a spin sensitivity $S_{\mu}^{1/2}\approx23\,\rm{\mu_B/Hz^{1/2}}$.
Concerning the suitability to applied magnetic fields, we successfully redesigned the layout of SQUID\,1 via FIB milling and implemented these findings into the design of SQUID\,2.
We demonstrated stable operation in a field range of $B\approx\pm50\,\rm{mT}$ with a marginal increase in white flux noise and spin sensitivity with $B$ ($S_{\Phi,\rm{w}}^{1/2}\leq250\,\rm{n \Phi_0/Hz^{1/2}}$ and $S_{\mu}^{1/2}\leq29\,\rm{\mu_B/Hz^{1/2}}$).
Moreover it was shown that SQUID\,2 can maintain high sensitivity in large fields up to $B=0.5\,\rm{T}$ with $S_{\Phi,\rm{w}}^{1/2}\approx680\,\rm{n \Phi_0/Hz^{1/2}}$ and $S_{\mu}^{1/2}\approx79\,\rm{\mu_B/Hz^{1/2}}$.
An obvious way to further decrease $S_{\Phi}^{1/2}$ and $S_{\mu}^{1/2}$ is to lower the SQUID inductance $L$, which can be done easily by decreasing the lateral distance between the JJs and by reducing the thickness of the $\rm{SiO_2}$ layer separating the top and bottom Nb layers.
In addition, the width of the Nb lines can be reduced further to increase $\phi_{\mu}$ and to extend the range of magnetic fields where the SQUID can be operated without vortices entering the wiring.
All in all, we consider a spin sensitivity down to a few $\rm{\mu_B/Hz^{1/2}}$, for a field range exceeding $100\,\rm{mT}$, to be achievable for this type of device.

J.~Nagel and T.~Schwarz gratefully acknowledge support by the Carl-Zeiss-Stiftung.
We thank K.~St\"{o}rr for assistance in device fabrication.
This work was funded by the DFG via the SFB TRR 21 and by the ERC via SOCATHES.

\end{document}